%% file: index.tex
\documentclass[conference]{IEEEtran}

\IEEEoverridecommandlockouts

\usepackage{tabularx,booktabs}
\usepackage[utf8]{inputenc}
\usepackage{float}
\usepackage{hyperref}
\usepackage{multirow}
\usepackage{cite}
\usepackage{amsmath,amssymb,amsfonts}
\usepackage{algorithmic}
\usepackage{graphicx}
\usepackage{textcomp}
\usepackage{xcolor}
\usepackage{xspace}
\usepackage{mathrsfs}

\def\BibTeX{{\rm B\kern-.05em{\sc i\kern-.025em b}\kern-.08em
    T\kern-.1667em\lower.7ex\hbox{E}\kern-.125emX}}

\newcommand{\etal}{\textit{et al.}\space}
\newcommand{\tool}{\textsc{MuCoRest}\xspace}

\begin{document}

\title{Reinforcement Learning-Based REST API Testing with Multi-Coverage}

\author{\IEEEauthorblockN{Tien-Quang Nguyen, Nghia-Hieu Cong, Ngoc-Minh Quach, Hieu Dinh Vo, and Son Nguyen$^*$\thanks{*Corresponding author.}}
\IEEEauthorblockA{\textit{Faculty of Information Technology} \\
\textit{University of Engineering and Technology, Vietnam National University, Hanoi, Vietnam} \\ \{20020116, 21020540, 20020261, hieuvd, sonnguyen\}@vnu.edu.vn} 
}

\maketitle

\input{0.abstract}
\input{1.intro}
\input{2.background}

\input{3.approach}
\input{4.eval_method}

\input{5.results}
\input{6.related_work}
\input{7.conclusion}



\section*{Acknowledgment}
This research is supported by Vietnam National Foundation for Science and Technology Development (NAFOSTED) under grant number 102.03-2023.14.

\bibliographystyle{IEEEtran}
\bibliography{ref}
\end{document}

%% file: 0.abstract.tex
\begin{abstract}



REST (Representational State Transfer) APIs have become integral for data communication and exchange due to their simplicity, scalability, and compatibility with web standards. However, ensuring REST APIs' reliability through rigorous testing poses significant challenges, given the complexities of operations, parameters, inputs, dependencies, and call sequences. In this paper, we introduce \tool, a novel Reinforcement Learning (RL)-based API testing approach that leverages Q-learning to maximize code coverage and output coverage, thereby improving bug discovery. By focusing on these proximate objectives rather than the abstract goal of maximizing failures, \tool effectively discovers critical code areas and diverse API behaviors. The experimental results on a benchmark of 10 services show that \tool significantly outperforms state-of-the-art API testing approaches by 11.6–261.1\% in the number of discovered API bugs. \tool can generate much fewer API calls to discover the same number of bugs compared to the other approaches. 
Furthermore, 12.17\%--64.09\% of the bugs discovered by the other techniques can also be found by \tool.


\end{abstract}

\begin{IEEEkeywords}
Automated RESTful API Testing, Reinforcement Learning for Testing.
\end{IEEEkeywords}

%% file: 1.intro.tex
\section{Introduction \label{introduction}}

In recent years, REST (Representational State Transfer) APIs have gained popularity for communicating and exchanging data due to their simplicity, scalability, and compatibility with web standards~\cite{richardson2013restful} \cite{patni2017pro}.  
Their design, which emphasizes reusability and interoperability, allows seamless integration across various components and systems. 
Consequently, rigorous REST API testing is imperative to ensure these APIs perform reliably, thereby maintaining high-quality interactions across all application clients.
Inherited from general software testing practices, one of the most important goals of API testing is discovering a diverse range of outputs/bugs. Identifying a wide variety of issues early in the development process helps API developers address potential problems before they affect application clients and end-users.


Achieving the above goal in REST API testing could be challenging because of the large numbers of operations, parameters, inputs, their dependencies/constraints, and the huge set of potential sequences of calling the API operations~\cite{empirical-study,arat,morest}.
Recently, Atlidakis \etal~\cite{restler} proposed RESTler, a fuzzing tool for automatically testing REST APIs, which generates stateful test cases by utilizing the dependencies between operations. Morest~\cite{morest} focuses on dynamically generating valid and meaningful sequences of API calls (aka \textit{call sequences}). 
Those approaches often struggle to explore API bugs due to the lack of consideration of the importance of operations and parameters to the overall goal (discovering a diverse range of outputs/bugs). 
Kim~\etal\cite{arat} introduce ARAT-RL, a reinforcement learning (RL)-based approach specifically targeting API bug discovery.
Particularly, ARAT-RL directly assigns a positive reward for \texttt{4xx} and \texttt{5xx} response status codes, indicating unsuccessful requests and API bugs during API call generation and execution. 
%
%
%
%
However, this \textit{ultimate objective} of maximizing the number of failures could be quite abstract and hard for RL agents to manage and learn to discover API bugs effectively.

In this paper, we introduce \tool, a novel Reinforcement Learning-based API testing approach. In \tool, the Reinforcement Learning (RL) agent pursues the \textit{proximate} and \textit{more manageable objectives}, maximizing the code coverage and output coverage in generating API calls to discover bugs instead of the abstract ultimate objective of maximizing the number of failures. 
Our intuition is that by combining these two immediate objectives, \tool could better discover and test the critical code areas implementing the API and explore a more diverse range of the APIs' behaviors, ultimately detecting more not-yet-discovered API bugs.



In \tool, we apply the Q-learning algorithm with the reward function specialized for the API testing task.
Particularly, the reward function encourages the agent to generate API calls that maximize API bug discoverability through the two immediate objectives of maximizing code coverage and output coverage. The code coverage is analyzed to determine the appropriate reward for each increment stage. Meanwhile, the reward for output coverage improvement is determined based on the uniqueness of the API calls' outputs. Additionally, the agent is rewarded once it generates an API call, which effectively reveals not-discovered-yet API bugs.


We conducted several experiments to evaluate \tool's performance on a benchmark, which is widely used by existing studies~\cite{empirical-study} and contains 10 services with 860K LOCs and about 200 operations. 
%
%
Our results show that \tool discovers \textbf{11.6--261.1\%} more API bugs than the state-of-the-art API testing approaches. Additionally, \tool can discover API bugs much faster than the other techniques. Among the API bugs found by the other approaches, \textbf{12.17--64.09\%} of them are also covered by \tool.

%

In brief, this paper makes the following contributions:
\begin{enumerate}
    \item \tool: A novel RL-based API testing approach which effectively combines the objectives of maximizing the code coverage and output coverage in discovering API bugs.

    \item An extensive evaluation showing the performance of \tool over the state-of-the-art API testing approaches in discovering API bugs.
\end{enumerate}



The rest of this paper is organized as follows. Section~\ref{background} describes the essential basic concepts. Our REST API testing approach is introduced in Section~\ref{approach}. 
After that, Section~\ref{eval_method} states our evaluation methodology. Section~\ref{sec:results} presents the experimental results following the introduced methodology and some threats to validity.
Section~\ref{relatedwork} provides the related work. Finally, Section~\ref{conclusion} concludes this paper.

%% file: 2.background.tex
\section{Background \label{background}}


\subsection{REST APIs}

REST (Representational State Transfer) APIs are web APIs that follow REST architectural style~\cite{RESTAPI}. They allow clients to interact with web services by sending HTTP \cite{http} \textit{requests} and receiving \textit{responses}. A request is sent to an API endpoint, a resource path that identifies the resource, along with a standard HTTP method (GET, POST, PUT, DELETE, and PATCH) that specifies the action to be performed on the resource. The combination of an endpoint and an HTTP method is called an \textit{operation}. 
%
An API call is a request made by a client to a server to perform a specific operation. For an API, the API parameters allow clients to specify the data in API calls they want to send to/retrieve from the server.

Each API call generates a response with a status code that indicates the outcome of the request. Successful responses usually have the status codes in the \texttt{2xx} range, such as \texttt{200} (OK) or \texttt{201} (Created). Client-side errors, indicated by \texttt{4xx} status codes such as \texttt{400} (Bad request) or \texttt{404} (Not Found), often result from invalid input parameters or unauthorized access attempts. Server-side errors, indicated by \texttt{5xx} status codes such as \texttt{500} (Internal Server Error), represent issues on the server that prevent it from fulfilling valid requests. Returning \texttt{4xx} and \texttt{5xx} from an API call with valid input parameters indicates a \textit{failures} caused by \textit{API bugs}.

\subsection{Reinforcement Learning and Q-learning}

Reinforcement Learning (RL) is a type of machine learning where an agent learns to achieve a goal by interacting with its environment~\cite{RL}. 
The agent makes decisions through a process of trial and error and receives feedback in the form of rewards or punishments based on the actions it takes, aiming to maximize cumulative rewards over time. 
This approach mimics the way humans and animals learn from experiences. 
Key RL concepts include the \textit{agent}, \textit{environment}, \textit{states}, \textit{actions}, and \textit{rewards}. 
The agent perceives the state of the environment and decides on an action to perform, which in turn affects the environment, transitioning it to a new state and yielding a reward. The cycle of observation, action, and reward is central to the RL process.
%

Q-learning is a specific type of RL algorithm that focuses on learning the value of action-state pairs, known as Q-values~\cite{Q-Learning}. Through Q-learning, a program dynamically refines its value estimates based on received rewards, progressively enhancing its ability to make decisions. These values are recorded in a Q-table. 
The algorithm operates by maintaining the Q-table, where each entry corresponds to a state-action pair and represents the expected cumulative reward for taking that action in that state. Initially, the Q-values are arbitrary, but through iterative updates based on the agent's experiences, the Q-values converge to reflect the true value of actions. The update rule in Q-learning uses the Bellman equation \cite{2018reinforcement}:
$$
Q(s,a) \leftarrow Q(s,a) + \alpha[r + \gamma \max_{a'} Q(s',a') - Q(s,a)]
$$
where $s$ is the current state, $a$ is the action taken, $r$ is the reward received, $s'$ is the next state, $\alpha$ is the learning rate, and $\gamma$ is the discount factor. This update rule ensures that the Q-value for a state-action pair gets closer to the true value of taking that action in that state, considering future rewards. 



%% file: 3.approach.tex
\section{\tool: A RL-based REST API Testing Approach with Multi-Coverage}
\label{approach}

This section describes, \tool, our novel RL-based API testing approach, which is designed to discover API bugs. In \tool, this long-term goal is pursued through two immediate objectives in discovering API bugs: maximizing the code coverage and output coverage.
We model the task of discovering API bugs as a RL task. Specifically, in \tool, the \textit{system under test} serves as the \textit{environment} and the \textit{agent} is responsible for generating API calls and executing the calls. The \textit{state} is expressed via current code coverage, output coverage, and other state variables, and the \textit{actions} include selecting operations and input parameters to form API calls. The \textit{reward} is determined by evaluating the code coverage, output coverage, and API calls' response.

\begin{figure}
	\includegraphics[width=1\columnwidth]{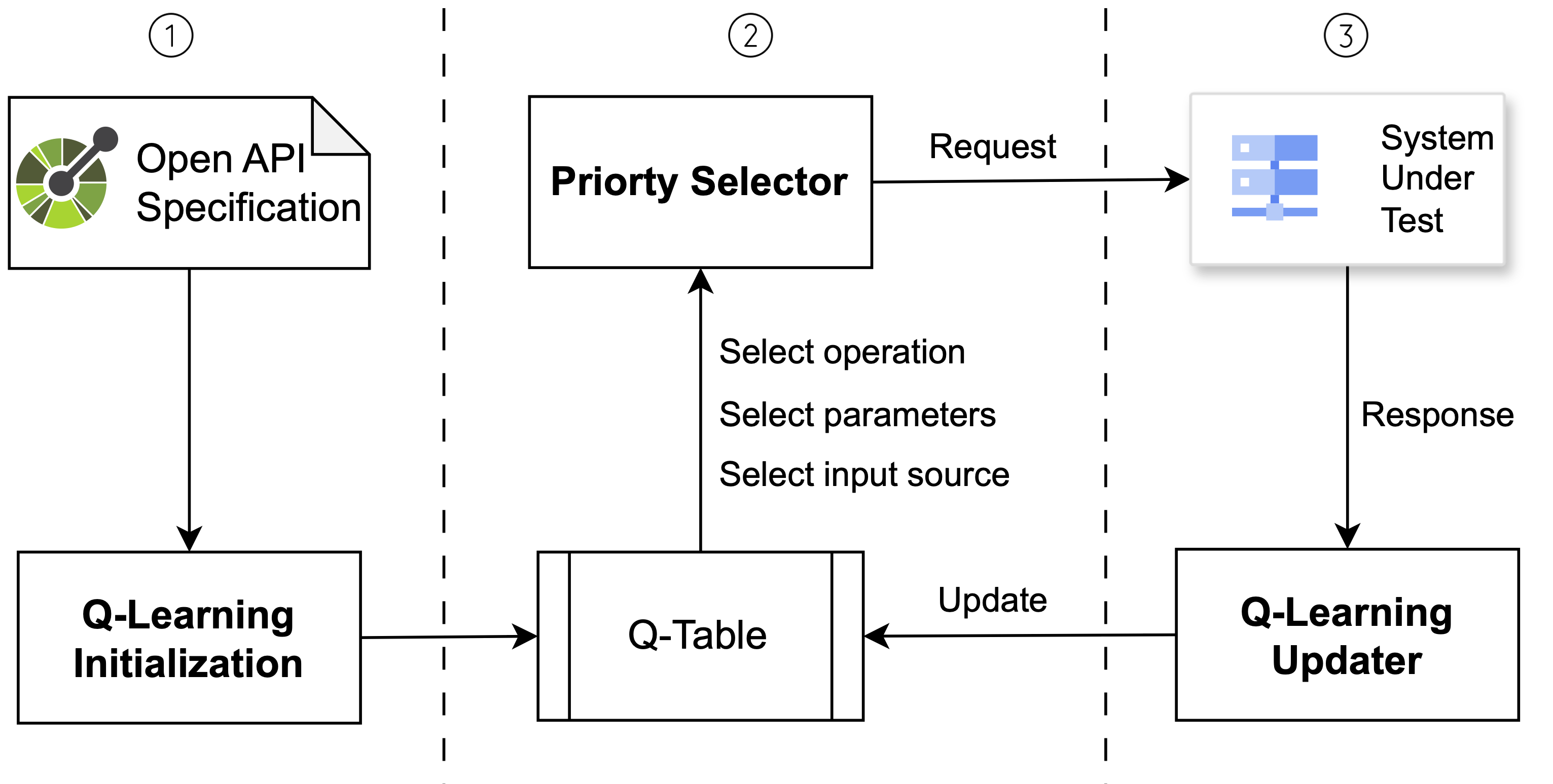}
	\centering
	\caption{Q-learning algorithm for RL-based API testing}
	\label{img:arat-rl-workflow}
\end{figure}

In this work, we reuse the basic components of the Q-learning algorithm for API testing in ARAT-RL~\cite{arat}. Figure \ref{img:arat-rl-workflow} shows an overview of the Q-learning algorithm, which consists of three main stages.
In the first stage, Q-Learning table initialization, the Q-table is initialized by analyzing parameter usage frequency in operations and prioritizing common parameters for early examination. 
%
In the second stage, test case inputs are created by selecting operation, parameters and their values. 
%
%
Finally, the algorithm updates the Q-table after test case execution based on the improvements of \textit{code coverage}, \textit{output coverage}, and \textit{API calls' response}.

\subsection{Code Coverage}

%
Essentially, improving code coverage could help the agent explore more diverse code areas implementing the APIs, increasing the likelihood of discovering more bugs in the APIs' implementation.
In \tool, we encourage the agent to explore untested areas of the API by tracking the code areas executed by each API call and rewarding for the calls that help explore not-yet-covered code areas. 
%
%
%
Particularly, \tool assesses code coverage gains by tracking the number of new lines of code executed for each API call. The rewards are given to the agent based on the proportion of the additional lines of code executed relative to the total source code. 
However, it is important to balance the significance of newly discovered code against the overall code coverage. In \tool, we model the increase of code coverage in two stages: the \textit{fast-growing} and the \textit{stabilizing}.

In the \textit{fast-growing} stage, as many code statements have not been executed yet by the API calls generated by \tool, it is still easy for the agent to generate API calls to execute new code statements, and the code coverage significantly increases. Thus, the reward for improving the coverage should be relatively small. 
In the \textit{stabilizing} stage, the possibility of generating calls to execute not-yet-covered code statements is greatly reduced. Consequently, it becomes much harder to improve the coverage, and the reward for improving code coverage should be relatively higher than that for the \textit{fast-growing} stage. 
%
Formally, the reward for improving code coverage of a generated API call, $c$, is calculated as follows:
%
%
%
%
\begin{equation}  
\label{eq:code-coverage}
r_{cc} (\mathscr{C}_{c}, \mathscr{C})=
  \begin{cases}
    0       & \quad \text{if } \mathscr{C}_{c} = \mathscr{C}\\
    R_{fg}       & \quad \text{if } \mathscr{C}_{c} - \mathscr{C} > 0 \text{ \& is \textit{fast-growing}} \\
    2 \times R_{fg}      & \quad \text{if } \mathscr{C}_{c} - \mathscr{C} > 0 \text{ \& is \textit{stabilizing}}
  \end{cases}
\end{equation}
where $\mathscr{C}$ and $\mathscr{C}_{c}$ are the code coverage before and after executing the API call $c$. In Formula~\ref{eq:code-coverage}, there is no reward if executing $c$ does not improve the code coverage. In the case of an improvement ($\mathscr{C}_{c} - \mathscr{C} > 0$), if it is greater than a predefined threshold (fast-growing), the reward is $R_{fg}$. Otherwise (stabilizing), the reward is $2 \times R_{fg}$. In this work, we consider the statement coverage and the predefined threshold as half of the average coverage improvement of the first 100 generated calls. 

In this work, we utilize Jacoco\footnote{https://www.jacoco.org/}, a widely used Java code coverage library, to collect accumulated code coverage ($\mathscr{C}_{c}$) for each API call. Note that \tool collects only the accumulated code coverage without considering the source code of services under test.

\subsection{Output Coverage}

Output coverage measures the diversity of outputs produced by systems under test. Intuitively, covering more diverse outputs could help the agent discover a larger set of systems' behaviors, improving the bug detection capability in both quantity and diversity. In \tool, the output coverage is improved by encouraging the agent to generate the API calls with unique outputs. 
The reward for improving output coverage is given based on the uniqueness of outputs produced by the generated API calls. 
%
%
When a new API call is generated, we assess the uniqueness of its output by comparing it with the outputs of the $H > 0$ closest historical calls that share the same operation and status code. Below is the formula used to estimate the reward for output coverage:




%
\begin{equation}  
\label{eq:output-coverage}
r_{oc} =
  \begin{cases}
    0   &  \text{if no resp. or \texttt{401}/\texttt{403} codes} \\
    R_{uniq.} \times (1 - \frac{2N}{H})    & \text{otherwise}
  \end{cases}
\end{equation}
where $N$ ($N \leq H$) represents the number of calls' responses among the $H$ closest historical calls that exactly match the evaluated response, and $R_{uniq.}$ is the predefined reward for the response's uniqueness. In Formula \ref{eq:output-coverage}, no reward is granted if the call returns  \textit{No response} or the status codes of \textit{401 (Unauthorized)} or \textit{403 (Forbidden)}.
Otherwise, the status codes such as \texttt{5xx}, other \texttt{4xx}, \texttt{3xx}, and \texttt{2xx} signify that some API functionality has been invoked. However, repeatedly receiving the same response does not contribute to effective testing, even if they are errors. Instead, comparing the body of the current response with the bodies of the previous $H$
 responses of the same type can be more insightful. Discovering a unique response body could reveal new issues, warranting a significant reward in the testing process.



\subsection{Bug Discoverability}


Discovering a larger number and wider range of API bugs causing failures is the agent's ultimate objective in \tool. Thus, the agent should be given more rewards once it can generate an API call to trigger a failure in the system under test. However, as a wider range of bugs/failures is preferred, the agent should be ``punished'' once it repeatedly generates calls that trigger the same bugs.
The reward for bug discoverability is calculated as follows:

\begin{equation}  
\label{eq:bug_discoverability}
r_{bd} =
  \begin{cases}
    R_{denied}             & \quad \text{if \texttt{401}/\texttt{403} status codes or no resp.} \\
    R_{invalid}             & \quad \text{if other \texttt{4xx} status codes} \\
    R_{success}             & \quad \text{if \texttt{2xx}/\texttt{3xx} status codes} \\
    R_{failure} \times p       & \quad \text{if \texttt{5xx} status codes} \\
  \end{cases}
\end{equation}
where $R_{denied}$, $R_{invalid}$, $R_{success}$, and $R_{failure}$ are the predefined rewards for the different cases of response status code.
It is recommended to reward the RL agent generously and assign a high value to the variable $R_{failure}$ when dealing with \texttt{5xx} status codes, as they indicate the discovery of a new bug.
To address the cases of repeated failures, a coefficient denoted as $p$ is used to prevent the testing process from getting stuck on a specific error.
Note that the repeated failures are the failures with identical error logs.
Intuitively, the more frequently failure is repeated, the lower $p$ correspondingly.
Inaccessible endpoints should not be encountered during testing, so a negative value for $R_{denied}$ is suggested.
Regarding other response status codes, they broaden the output coverage but do not advance the identification of failures. Therefore, a small positive value is recommended for $R_{invalid}$ and $R_{success}$.

\textit{Overall}, the formula for the total reward assigned to the agent per API call in \tool is:
\begin{equation}
\label{eq:rewards}
    r_{total} = r_{cc} + r_{oc} + r_{bd}
\end{equation}

%% file: 4.eval_method.tex
\section{Evaluation Methodology}
\label{eval_method}

To evaluate our API testing approach, we seek to answer the following research questions:

\textbf{RQ1: \textit{Accuracy and Comparison.}}  How does \tool perform in discovering API bugs  compared to state-of-the-art API testing approaches~\cite{arat,evomaster,morest}?



\textbf{RQ2: \textit{Objective Analysis}.} How do the immediate objectives based on code coverage and output coverage contribute to the overall performance of \tool?

\textbf{RQ3: \textit{Time Complexity}.} What is \tool’s running time?

\subsection{Dataset}

In this work, we evaluate the API testing approaches with the same benchmark used in the existing work~\cite{arat,empirical-study}, which contains ten open-source RESTful services with about 200 operations and 860K lines of code. Specifically, they provide OpenAPI specifications, which can be compiled and operated well and do not depend on external services with limited requests. This ensures that the performance of the testing tool is not affected by external factors. 

\subsection{Evaluation Procedure}
We compare \tool against the state-of-the-art API testing approaches, including:
\begin{enumerate}

    \item \textbf{ARAT-RL}~\cite{arat}: A automated REST API testing tool that applies Reinforcement Learning to test operations based on priority criteria to enhance testing performance.
     
    \item \textbf{Morest}~\cite{morest}: A stateful API testing tool that utilizes the generation of RESTful Service Property Graph (RPG) with extracted dependencies between APIs.
    
   
   \item \textbf{EvoMaster}~\cite{evomaster}: An open-source tool that can automatically generate system-level test cases using evolutionary algorithms. 
   The tool has two modes: white box and black box. This work uses the black-box mode as the baseline for our tool comparison.

\end{enumerate}

We apply the same evaluation procedure as in the existing studies~\cite{arat,empirical-study}. Each approach is executed with a time limit of 20K API calls. The experiments were repeated ten times. 

For \textbf{RQ2: Objective Analysis}, we examined the impact of incorporating each coverage information in testing on \tool's performance. 

    

For all experiments, we evaluate \tool and the baselines based on \textbf{the number of discovered API bugs}. We focused on identifying \textit{\textbf{unique}} internal server bugs with a response status code of \texttt{500} whose log message is unique.

%% file: 5.results.tex
\section{Experimental Results}
\label{sec:results}
\subsection{Accuracy and Comparison (RQ1)}

\begin{figure}
    \centering
    \includegraphics[width=1.0\columnwidth]{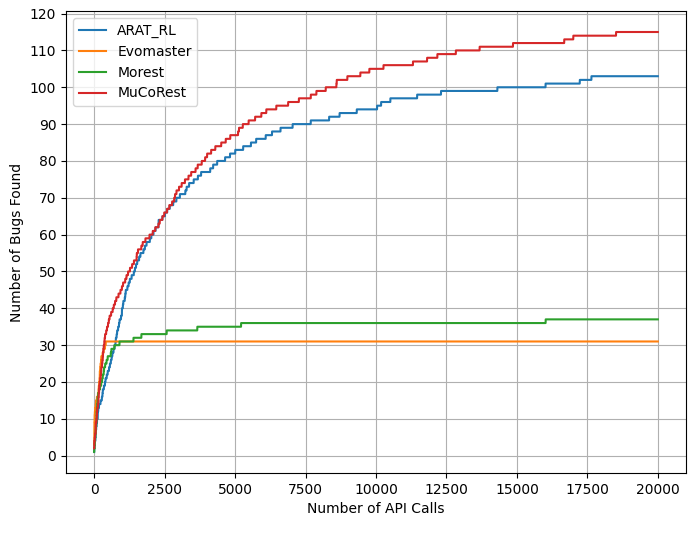}
    \caption{Relationship between API calls and discovered bugs}
    \label{figure:Bug-rquest}
\end{figure}

Table \ref{tbl:bug-detection} shows the number of discovered bugs by each approach across ten runs for the REST services in our benchmark. Note that this cumulative count might include unique bugs detected in multiple runs of the same fault in different runs. 
As seen, \tool demonstrated the most effective capability in discovering API bugs, significantly outperforming the state-of-the-art REST API testing approaches with a total of 1,152 found API bugs, which is 11.6\% to more than 261.1\% better than the corresponding figures of the others. 
Meanwhile, ARAT-RL, Morest, and EvoMaster discovered 1,032 bugs, 372 bugs, and 319 bugs, respectively. 
%
%
Especially, for ``Language Tool'' and ``Person Controller'', \tool effectively discovered 91 and 1,001 faults, respectively. These figures are significantly larger than those for the other services because ``Language Tool'' and ``Person Controller'' possess larger parameter sets. This enables testing with more diverse parameter combinations, leading to greater output variability and a higher likelihood of fault detection.
Additionally, as seen in Figure \ref{figure:Bug-rquest}, \tool tends to cost fewer API calls to discover API bugs compared to the other approaches. Particularly, ARAT-RL needed to generate 7,500 API calls to discover 90 bugs, while \tool generated only 5,000 calls to discover the same number of API bugs.

\input{tabs/tab_discovered_bugs}

\begin{figure}
    \centering
    \includegraphics[width=1\columnwidth]{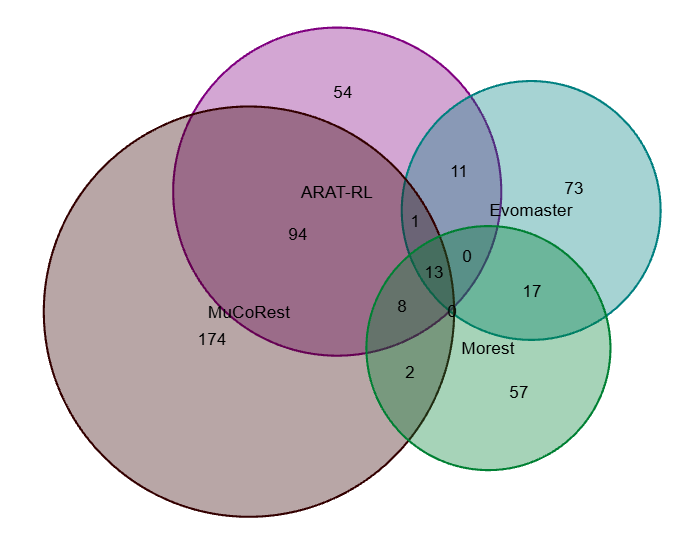}
    \caption{The sets of bugs discovered by \tool and the other API testing approaches}
    \label{figure:venn}
\end{figure}

In this experiment, we further investigate the API bugs discovered by \tool and the other studied approaches. Figure~\ref{figure:venn} illustrates the sets of API bugs discovered by each approach over 10 runs. 
\tool discovered 174 bugs, which cannot be found by any of the other approaches. This set is by far larger than the corresponding sets of the others. Additionally, \tool discovered 12.17\%--64.09\% the bugs found by the others. For example, among 181 bugs found by ARAT-RL, 116 bugs could be discovered by \tool. This further confirms the effectiveness of \tool in discovering API bugs compared to the state-of-the-art REST API testing approaches.

\subsection{Objective Analysis (RQ2)}
\begin{figure}

\centering{
\includegraphics[width=1.1\columnwidth]{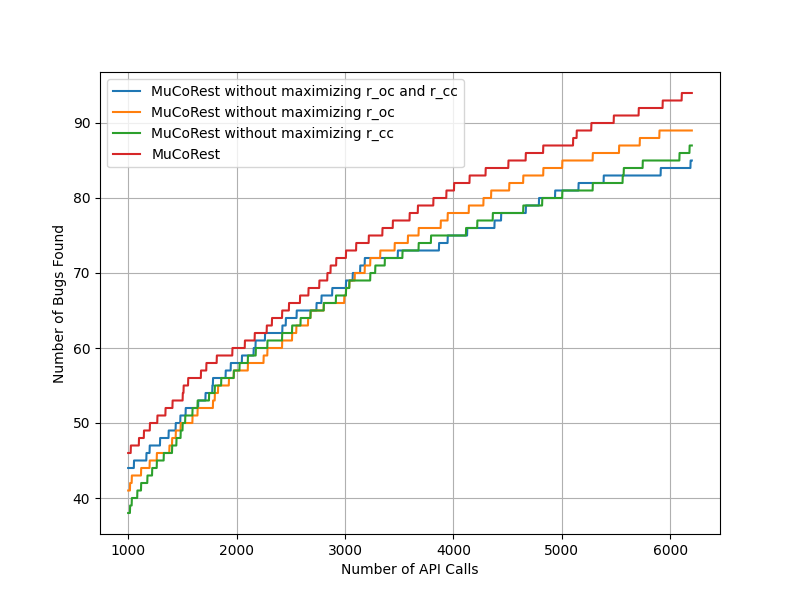}
\caption{Objective analysis}
\label{figure:Ablation-study}

}
\end{figure}
We additionally conducted experiments to investigate the impact of the main components of \tool on its performance. We compared the performance of \tool to three variants based on its objectives (Formula \ref{eq:rewards}): (1) \tool without maximizing code coverage, (2) \tool without maximizing output coverage, and (3) \tool without maximizing both output coverage and code coverage. 
As shown in Figure~\ref{figure:Ablation-study}, \tool performed most effectively when considering both immediate objectives along with the ultimate objective. Indeed, \tool discovered 4 and 2 more bugs when maximizing code coverage and output coverage compared to \tool without considering both output coverage and code coverage.
Meanwhile, maximizing both immediate objectives helps \tool improve the number of found bugs in discovering API bugs by 11\%. These results demonstrate the effectiveness of our consideration of the immediate objectives in \tool.


\subsection{Time Complexity (RQ3)}
In this work, all experiments were conducted on a personal machine running Ubuntu 20.04 with an Intel Core i5-8400 2.8GHz (6 cores) and 16GB RAM. In
\tool, it takes approximately 0.3 to 0.5 seconds to generate an API call to services.

\subsection{Threats to Validity}

The main threats to the validity of our work consist of internal, construct, and external threats.

\textbf{Threats to internal validity} include the impact of the method used to collect code coverage. To reduce this threat, we use the widely-used Java code coverage tool Jacoco.
Another threat lies in the correctness of the implementation of our approach. To reduce such a threat, we carefully reviewed our code and made it public~\cite{website} so that other researchers can double-check and reproduce our experiments.

\textbf{Threats to construct validity} relate to the suitability of our evaluation method. We used the number of unique discovered bugs to thoroughly examine the effectiveness of the approaches. 
In addition, a threat may come from the application of the baselines. To mitigate this threat, we directly obtained the original source code from their GitHub repositories or replicated exactly their description and used the same hyper-parameters specified in the paper~\cite{arat,evomaster,morest}.


\textbf{Threats to external validity} mainly lie in the benchmark services. The set of ten Java services could not generally reflect every service. Furthermore, as these services were run locally, they could behave differently compared to real-world services. In our future work, we plan to conduct more experiments to validate our results with a larger set of services.

%% file: tabs/tab_discovered_bugs.tex
\begin{table}
\centering
\caption{TOTAL BUG DETECTED BY TOOLS ACROSS 10 RUNS}
\label{tbl:bug-detection}
\begin{tabular}{l|rrrr}
\hline
Service           & \multicolumn{1}{l}{EvoMaster} & \multicolumn{1}{c}{Morest} & \multicolumn{1}{c}{ARAT-RL} & \multicolumn{1}{c}{\tool} \\ \hline
Feature Service   & 10                            & 10                         & 10                          & 10                                       \\
Language Tool     & 48                            & 0                          & 78                          & \textbf{91}                              \\
NCS               & 0                             & 0                          & 0                           & 0                                        \\
Rest Countries    & 10                            & 10                         & 10                          & 10                                       \\
SCS               & 0                             & 0                          & 0                           & 0                                        \\
Genome Nexus      & 0                             & 6                          & 10                          & 10                                       \\
Person Controller & 221                           & 316                        & 894                         & \textbf{1,001}                            \\
User Management   & 10                            & 10                         & 10                          & 10                                       \\
Market Service    & 10                            & 10                         & 10                          & 10                                       \\
Project Tracking  & 10                            & 10                         & 10                          & 10                                       \\ \hline
Total             & 319                           & 372                        & 1,032                        & \textbf{1,152}                            \\ \hline
\end{tabular}%
\end{table}

%% file: 6.related_work.tex
\section{Related Work}
 \label{relatedwork}

\tool relates to the automated REST API testing studies. 
%
EvoMaster~\cite{evomaster} uses an evolutionary algorithm to generate test cases and discover operational dependencies.
RESTler~\cite{restler} generates API call sequences by inferring producer-consumer dependencies from specifications and analyzing dynamic feedback from prior test executions.
Meanwhile, Morest~\cite{morest} generates a RESTful-service Property Graph (RPG) to outline detailed API dependencies and refine captured dependencies dynamically during testing.
%
Recently, Kim~\etal propose ARAT-RL~\cite{arat} which addresses the challenge of navigating a vast space of API operations, their possible execution sequences, dependencies between parameters, and constraints on input values.
ARAT-RL also employs a dynamic testing strategy driven by a reinforcement learning algorithm to prioritize and explore the various API operations and parameters.
Different from prior studies, \tool targets more manageable goals, maximizing the code coverage and output coverage, rather than focusing on the abstract ultimate goal of maximizing the number of failures.
However, as analyzed in Section~\ref{sec:results}, \tool and the other approaches could complement to each others to discover a wider range of API bugs.


Several ML-based studies have been proposed for specific software engineering tasks, including code suggestion/completion~\cite{icse20, autosc,arist,manh2022novel}, code summarization~\cite{patchexplainer,mastropaolo2021studying}, pull request description generation~\cite{liu2019automatic}, fuzz testing\cite{godefroid2017learn}, code-text translation~\cite{ase22}, and bug/vulnerability detection~\cite{oppsla19, codejit}.

%% file: 7.conclusion.tex
\section{Conclusion }
\label{conclusion}
In this paper, we introduce \tool, a novel approach based on Reinforcement Learning for automated detection of vulnerabilities in RESTful APIs.
By using Q-Learning and multiple test coverages during the testing process, \tool allows the RL agent to test critical code areas that implement the API and explore a wider range of the APIs’ behaviors.
These features address issues in guiding the RL agent to achieve the abstract ultimate goal of maximizing the number of failures by focusing on two more immediate objectives: code coverage and output coverage. 
The experimental results on a benchmark of 10 services show that \tool significantly outperforms state-of-the-art API testing approaches by 11.6–261.1\% in discovering API bugs. Furthermore, 12.17\%--64.09\% of the bugs discovered by the other techniques can also be found by \tool.
